\documentclass{article}

\usepackage{PRIMEarxiv}

\usepackage[utf8]{inputenc} 
\usepackage[T1]{fontenc}    
\usepackage{hyperref}       
\usepackage{url}            
\usepackage{booktabs}       
\usepackage{amsfonts}       
\usepackage{amsmath}
\usepackage{amssymb}
\usepackage{nicefrac}       
\usepackage{microtype}      
\usepackage{graphicx}
\usepackage{fancyhdr}       
\usepackage{multirow}
\usepackage{subcaption}
\usepackage[table]{xcolor}
\usepackage{lipsum}
\graphicspath{{media/}}

\begin{document}

\title{Enhancing Eye Movement Biometrics for User Authentication via Continuous Gaze Offset Score Fusion
\thanks{\textit{\underline{Citation}}:
\textbf{Aziz, Raju, Komogortsev. Enhancing Eye Movement Biometrics for User Authentication via Continuous Gaze Offset Score Fusion. arXiv preprint, 2026.}}
}

\author{
    Hashim Aziz \\
    Texas State University \\
    San Marcos, Texas, USA \\
    \texttt{uwx6@txstate.edu} \\
    \And
    Mehedi Hasan Raju \\
    Texas State University \\
    San Marcos, Texas, USA \\
    \texttt{m.raju@txstate.edu} \\
    \And
    Oleg V. Komogortsev \\
    Texas State University \\
    San Marcos, Texas, USA \\
    \texttt{ok@txstate.edu} \\
}

\maketitle

\begin{abstract}
Eye movement biometrics (EMB) use subject-specific gaze dynamics for user authentication and identification. 
Recent deep learning-based EMB systems achieve strong performance by modeling temporal eye movement behavior.
However, these systems typically overlook continuous gaze offset, despite prior evidence that it contains user-discriminative information.
This work examines whether continuous gaze offset can improve biometric performance when combined with existing biometric features.
We evaluate linear and nonlinear fusion methods on two publicly available datasets, collected via the lab-grade eye tracker and virtual reality headset across multiple tasks and observation durations.
Results indicate that fusion offers performance benefits on both datasets, particularly when using nonlinear fusion. 
Additionally, fusing biometric information across multiple tasks further improves authentication performance. 
These findings support the hypothesis that continuous gaze offset may serve as useful auxiliary information under conditions of degraded or noisy eye tracking.
\end{abstract}


\section{Introduction}

Eye movement biometrics (EMB) is the process of extracting and comparing the unique characteristics of an individual's oculomotor behavior to identify them \cite{Kasprowski2004BioAW}. EMB has emerged as a promising biometric modality that utilizes physiological and cognitive gaze dynamics for a robust, spoof-resistant user authentication \cite{RIGAS2017129, raju2022iris, rigas2015}. 
Furthermore, since eye movements are continuously generated during visual tasks, EMB offers the potential for continuous, passive user authentication \cite{RIGAS2017129}. 

Recent trends in EMB research have shifted toward deep learning (DL), with state-of-the-art frameworks such as Eye Know You Too (EKYT) demonstrating strong authentication performance using gaze signals from both high-quality lab-grade eye trackers \cite{Lohr2022TIFS} and consumer virtual reality (VR) headsets \cite{raju2025spw, lohr2024establishing}.

However, a critical research gap remains: these purely temporal architectures operate exclusively on dynamic eye movement patterns and do not explicitly incorporate continuous gaze offset information — the systematic, subject-specific spatial error between intended and actual gaze position throughout the whole recording scenario. 
This means that physical stimulus-driven targeting data is implicitly discarded, which may limit the robustness of temporal models in challenging environments where signal quality is degraded, such as VR \cite{lohr2024establishing}.  

The motivation for this work stems from evidence that physical gaze targeting behavior, specifically calibration patterns and spatial gaze error, carries highly distinctive, subject-specific signatures capable of user identification above chance levels \cite{Kasprowski2020ETRA}. Because these spatial offset patterns are repeatable and user-discriminative, they capture traits that are fundamentally different from temporal gaze velocity \cite{Rigas2014Spatial}.
This complementarity raises a natural research question:

\textit{Can score-level fusion of continuous spatial gaze offset with DL-based embeddings improve biometric authentication performance, particularly under degraded signal conditions, without requiring architectural changes to existing models?}

The contribution of this paper is a rigorous evaluation of score-level fusion between the state-of-the-art temporal EKYT model and explicit continuous gaze offset measurements using stimulus-driven target data. 
To address our research question, we implement and evaluate multiple biometric fusion strategies - ranging from linear weighted combinations to nonlinear tree-based ensemble classifiers. 
We conduct experiments across two datasets with significantly different acquisition conditions: a high-quality lab-grade eye tracker (GazeBase) \cite{Griffith2021GazeBase} and a lower signal-quality VR headset (GazeBaseVR) \cite{lohr2023gazebasevr}.

By adhering to strict subject-disjoint cross-validation protocols and assessing performance across multiple tasks and observation durations, our results provide empirical evidence that explicitly fusing continuous gaze offset measurements with temporal dynamics yields consistent authentication improvements. Furthermore, we show that combining biometric information across tasks provides additional performance gains, which shows the value of task diversity when extracting user-discriminative information. 

\section{Prior Work}

\subsection{Continuous Gaze Offset and Calibration Based Identification}
Early work in EMB has demonstrated that continuous gaze offset and calibration behavior can carry user-discriminative information \cite{Kasprowski2020ETRA}.
In particular, Kasprowski proposed identifying users by comparing uncalibrated eye tracker outputs against known target locations during a calibration-like task. 
By computing spatial error between estimated gaze points and ground-truth targets, each trial could be represented by a compact spatial signature \cite{Kasprowski2020ETRA}. 
These results provided early empirical evidence that continuous gaze offset information alone can encode a biometric signal.

Expanding on the biometric utility of gaze offset distributions, Rigas and Komogortsev proposed a framework that utilizes Fixation Density Maps (FDMs) to authenticate users based on their unique spatial attention patterns \cite{Rigas2014Spatial}. 
By projecting time-sampled eye movement signals into the spatial domain, they constructed probabilistic representations of user-specific gaze behavior. 
These templates were then compared using dissimilarity metrics, achieving an equal error rate (EER) as low as 10.8\%. 
This spatial project also demonstrated robustness at frequencies as low as 30 Hz \cite{Rigas2014Spatial}. 
This supports the idea that spatial patterns remain discriminative even under degraded conditions, making spatial features critical for more challenging acquisition environments. 

\subsection{Eye Movement Biometrics}
Early EMB research demonstrated that eye movement patterns exhibit person-specific characteristics and can be used for human identification when users are exposed to controlled visual stimuli \cite{Kasprowski2004BioAW}. 
These early systems typically relied on explicit detection of eye movement events and hand-crafted features derived from oculomotor behavior.

Building on this foundation, exploration of specific tasks, such as reading, to extract more complex oculomotor features started. 
For instance, Holland and Komogortsev \cite{KomogortsevIJCB2011} investigated biometric identification by analyzing scanpaths generated during reading tasks. 
By extracting a number of explicit eye movement metrics such as average fixation duration and saccade amplitude, they demonstrated that aggregated characteristics could uniquely identify individuals. Their approach utilized a weighted mean technique to consolidate 15 distinct metrics into a single similarity score. 
This early work established the viability of aggregating explicitly engineered oculomotor features for authentication. 
While modern systems have increasingly integrated end-to-end DL to model complex temporal dynamics, the foundational principles of feature extraction and fusion remain relevant for capturing discriminatory information. 

More recent work has shifted toward end-to-end DL approaches that learn discriminative representations directly from raw or minimally processed eye movement signals  \cite{Jia2018, lohrTBIOM, deepeyedentificationlive, abdelwahab2022deep}. 
Among these, EKYT \cite{Lohr2022TIFS} is a state-of-the-art model that employs a DenseNet-based architecture to learn fixed-dimensional embeddings from eye-movement velocity signals. 
EKYT demonstrated strong authentication performance on a high eye-tracking quality dataset and showed robustness across tasks, test-retest intervals, and degraded sampling rates.

\subsection{Multi-Biometric Score-Level Fusion}

The fusion of independent biometric modalities is a well-established practice for improving authentication reliability and addressing the limitations of single-modality systems \cite{Ross2006HandbookOM, periocular}. 
According to standardized multi-biometric frameworks, score-level fusion, in which the independent outputs of different matchers are consolidated into a single vector, provides an effective mechanism for distinguishing genuine from impostor authentication attempts \cite{Ross2006HandbookOM}. 

Rather than relying on simple arithmetic rules (e.g., the sum or max rules), classification schemes feed the score vectors into machine learning operations to learn complex non-linear decision boundaries \cite{Modak2018Fusion}. 
Prior research has demonstrated the strong efficacy of Random Forest algorithms in classifying multi-dimensional score vectors, establishing tree-based classifiers as a highly robust approach for biometric fusion \cite{Modak2018Fusion}. 

Evaluating these multi-biometric systems requires adherence to standardized protocols to ensure reproducibility and generalizability. 
Best practice dictates that classifier-based fusion models must be evaluated using K-fold cross-validation on subject-disjoint datasets to prevent identity leakage. 
Since biometric datasets are inherently imbalanced between impostor and genuine combinations, fusion classifiers must systematically address class imbalance during training \cite{Ross2006HandbookOM}. 
The methodology proposed in this work is rooted in these established standards to ensure reliable and repeatable results.

\section{Methodology}

\begin{figure*}[t]
\centering
\includegraphics[width=0.95\linewidth]{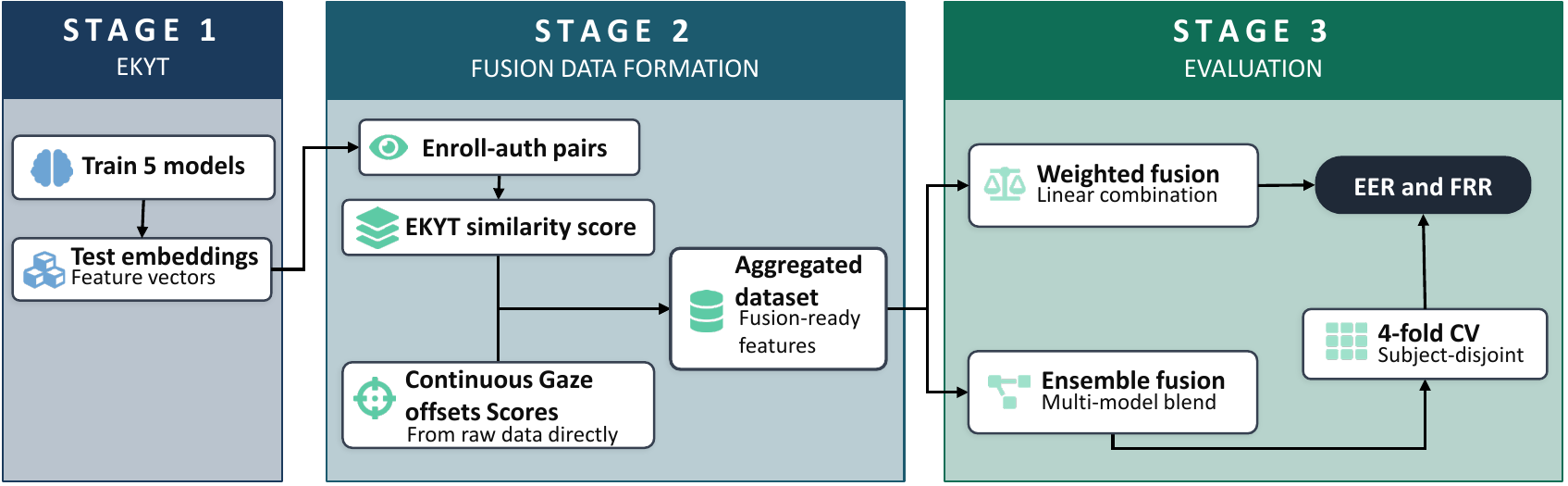}
\caption{Overview of the proposed eye movement biometric fusion pipeline, consisting of EKYT feature extraction (Stage 1), fusion dataset formation (Stage 2), and multi-modal performance evaluation using linear and non-linear ensemble methods (Stage 3).}
\label{fig:pipeline}
\end{figure*}

The methodology consists of three distinct steps: (1) EKYT model training and test embedding generation, (2) fusion dataset formation, and (3) fusion and final evaluation.
The proposed approach, illustrated in Figure \ref{fig:pipeline}, builds directly on the EKYT framework, with no architectural modifications, and introduces continuous gaze offset as an auxiliary signal fused with similarity scores calculated from EKYT learned embeddings.

\subsection{Datasets}

Experiments were conducted using two publicly available eye-tracking datasets: GazeBase and GazeBaseVR. 

GazeBase \cite{Griffith2021GazeBase} is a publicly available dataset of 12,334 monocular (left eye) recordings from 322 college-aged participants collected over three years across nine rounds.
Eye movements were recorded using an EyeLink 1000 system at 1000~Hz, capturing horizontal and vertical eye positions in degrees of visual angle (dva). 
Participants performed seven tasks: random saccades (RAN), reading (TEX), fixation (FXS), horizontal saccades (HSS), two video-viewing tasks (VD1, VD2), and a video-gaming task (BLG). 
Each round included two sessions per subject, separated by approximately 20 minutes. 
Further details are provided in \cite{Griffith2021GazeBase}.

GazeBaseVR (GBVR) \cite{lohr2023gazebasevr} is a GazeBase-inspired dataset collected with an ET-enabled VR headset (HTC SMI VIVE), containing 5,020 binocular recordings from 407 college-aged participants gathered over 26 months in three rounds. 
Eye movements were recorded at 250~Hz, capturing horizontal and vertical gaze positions of both eyes in dva. 
Participants completed five tasks: vergence (VRG), horizontal smooth pursuit (PUR), reading (TEX), video viewing (VD), and random saccades (RAN). 
Detailed data collection procedures are available in \cite{lohr2023gazebasevr}.

From each dataset, only the RAN and TEX tasks were used throughout all experiments. 
These two were chosen because they are both included in each dataset, and the RAN task includes the target information needed to calculate continuous gaze offset.

\subsection{Data Preprocessing}

Prior to training the EKYT model, raw gaze coordinates were preprocessed to extract continuous velocity signals.

First, the positional data was differentiated to obtain angular velocity in degrees per second using a Savitzky-Golay filter \cite{Savitzky1964}. 
A window length of 7 and a polynomial order of 2 were chosen to calculate the first derivative \cite{raju2023filtering}. 
This specific filter was chosen to smooth the signal and mitigate noise during the derivation process. 

After differentiation, the continuous velocity sequences were segmented into non-overlapping, fixed-length subsequences of 5 seconds. 
Finally, Z-score normalization was applied to the velocity subsequences where the mean and standard deviation are calculated using strictly the training set to prevent data leakage and standardize the validation and testing sets \cite{Lohr2022TIFS}.

\subsection{EKYT Model Architecture}

The EKYT architecture was selected for its state-of-the-art biometric performance \cite{Lohr2022TIFS}. 
Discrepancies between our baseline results and those in the original study arise from our evaluation protocol; specifically, we aggregate test embeddings across five independent models, each with a unique test set, to ensure a comprehensive and unbiased evaluation of the entire subject pool.

EKYT is a DenseNet-based architecture that achieves state-of-the-art EMB performance in an authentication situation using eye movement velocity signals \cite{Lohr2022TIFS}. It has demonstrated exceptional performance across multiple large-scale datasets. On high-quality laboratory data recorded at 1000 Hz (GazeBase), EKYT achieves a 0.58\% EER using 60 seconds of data from a reading task \cite{Lohr2022TIFS}. On noisier VR based data recorded at 250 Hz (GazeBaseVR), EKYT achieves an EER of 1.67\% using a data from a reading task in a binocular configuration \cite{raju2025spw}. Most notably, on a very large dataset of over 9,000 subjects recorded at 72 Hz (GazePro), EKYT achieved an EER of 3.73\% in a binocular configuration using data from a random saccade task \cite{lohr2024establishing}.

EKYT is a highly efficient architecture with about 123,000 learnable parameters, making it ideal for real-world deployments in resource-constrained environments. 
The network architecture has 8 convolutional layers where the feature maps created by each convolution layer are concatenated with all preceding feature maps. 
A 128-dimensional embedding of the input sequence is then created using the final set of concatenated feature maps after they have been flattened and sent through a global average pooling layer and fed into a fully connected layer. 
For more details about the network architecture, we refer the reader to \cite{Lohr2022TIFS}.

Each recording was divided into fixed-length subsequences of 5-seconds. Experiments varied the number of subsequences $n_{\text{seq}}$ used per evaluation, corresponding to total durations ranging from 5 to 40 seconds of eye-tracking data.

Since the EKYT pipeline was used without modification to the architecture, all improvements or degradations in performance therefore arise solely from the introduction of continuous gaze offset scoring and fusion strategies, rather than from changes to the underlying biometric model.

\subsection{Continuous Gaze Offset Scoring}
Continuous gaze offset captures how accurately an eye tracker can estimate a subject’s gaze based on their intended target \cite{raju2025ETRA}. 
In this experiment, it was used as a method of comparison between two subjects, and the resulting similarity score was fused with the EKYT similarity score with the goal of improving biometric performance. 

Continuous gaze offset was computed using an angular offset formulation applied to gaze samples from the RAN task. 
For each gaze sample, the angular distance between the estimated gaze position and the known stimulus target location was computed. 
This calculation was performed per row of each recording file, allowing continuous gaze offset statistics to be aggregated across time and across subsequences for use in fusion. 
Continuous gaze offset is calculated according to the following equation:

\begin{equation}
\theta =  \frac{180}{\pi} \arccos\left( \frac{G \cdot T}{\|G\| \|T\|} \right) \tag{3}
\end{equation}

where G is the centroid of the gaze-position vector, T is the target-position vector, and $\theta$ is the angular distance in degrees of visual angle \cite{raju2025ETRA}. 

After computing continuous gaze offset, fixations were identified using the dispersion identification threshold (IDT) algorithm \cite{2000Salvucci, 2009Blignaut}. 
Once fixations were identified, aggregated statistical information was extracted for each recording while filtering out any data where a fixation was not detected. 
A Euclidean distance-based similarity method was used to produce a similarity score for the authentication and enrollment pair. 
To translate Euclidean distance to a similarity score bound to the range [0,1], the following equation is used:
\begin{equation}
    s = \frac{1}{1 + d}
\end{equation}
where d is the calculated Euclidean distance between the two subjects. 

The same subsequences used for EKYT similarity scoring were also used for computing continuous gaze offset statistics, ensuring temporal alignment between biometric similarity and continuous gaze offset measurements.

\subsection{Fusion Techniques}

After the EKYT similarity scores ($S_\text{EKYT}$) and the continuous gaze offset similarity scores ($S_\text{Spatial}$) are computed, the scores are fused.
Score level fusion is evaluated using 4 methods: (1) tree-based ensemble fusion, (2) weighted fusion, (3) cross-task fusion, and (4) triple fusion. 
When fusing the EKYT TEX score with the continuous gaze offset score, the RAN task recording for the same subject, round, and session is used to calculate the continuous gaze offset score.

\subsubsection{Tree-Based Ensemble Fusion} 

Tree-based ensemble fusion was chosen because it provides a more robust method of fusion due to it being able to capture non-linear interactions between scores. 
The tree-based fusion model operates on an explicitly feature-engineered feature set derived from the EKYT similarity score and the continuous gaze offset similarity score. 
In addition to the raw scores, nonlinear interaction features are constructed to capture basic second-order relationships between the two scores. 
These engineered features are provided as input to the classifier, which further models complex decision boundaries. 
The fusion task is framed as a binary classification problem (genuine vs impostor) where the output probability is interpreted as the new fused similarity score. Separate models were trained for each task (RAN, TEX) and each subsequence length.

\subsubsection{Weighted Fusion}

Weighted fusion combines EKYT similarity scores and continuous gaze offset scores using a linear combination:
\begin{equation}
    S=\alpha * S_\text{EKYT}+(1-\alpha)*S_\text{Spatial}
\end{equation}

where the weighting parameter ($\alpha$) was evaluated over the range 0.5 – 1.00, in increments of 0.01 for each combination of task and duration of the recording.
The $\alpha$ value that yielded the best EER was kept and recorded along with the EER it produced. 
It is to be noted that in some cases, the optimal $\alpha$ value is 1.00, meaning the fused score relies entirely on one modality. 
This indicates that linear fusion provided no improvement over the individual system for that particular combination.

\subsubsection{Cross-task fusion}

Cross-task fusion was evaluated by fusing EKYT similarity scores computed from the RAN and TEX tasks. 
This fusion method doesn't incorporate any extra continuous gaze offset similarity information. 
This was tested to explore whether complementary biometric information captured across tasks can improve biometric performance. 
The EKYT similarity scores and non-linear interaction features between the scores were treated as inputs to a random forest fusion model. The fusion model was framed as a binary classification problem distinguishing genuine and impostor pairs. 

\subsubsection{Triple Fusion}

Triple fusion was evaluated by fusing the EKYT similarity score from the RAN and TEX tasks with the continuous gaze offset similarity score. Triple fusion uses the same engineered feature set that captures non-linear interaction features between the three scores as inputs to the same random forest fusion model that cross-task fusion uses.

\subsection{Training Procedure}
\subsubsection{Train and Test Split} 

All available rounds were used during EKYT training following the standard EKYT pipeline. While multiple tasks are included in both datasets, this study focuses evaluations on the RAN and TEX tasks, as these are common across both GazeBase and GazeBaseVR, and continuous gaze offset can be computed using the RAN task. 

No additional preprocessing beyond that required by EKYT was introduced. EKYT data modules handle the serialization of the raw data into .pkl files. 
For continuous gaze offset scoring, raw gaze and target coordinates were used directly to compute angular offsets after filtering for fixations using the IDT algorithm. 

\subsubsection{EKYT Training} 
The EKYT model was trained from scratch using all data from rounds 1-5 except for the BLG task. Subjects in round 6 were treated as the held-out subjects and were not used for training or validation. The entire EKYT pipeline was trained independently 5 times. 
For each of these 5 runs, a unique non-overlapping subset of subjects was completely held out as a test set. 
During training, the non-held-out subjects were further divided into 4 non-overlapping folds following standard EKYT cross-validation. In training, we used multi-similarity (MS) loss implemented via PyTorch. 
Training the pipeline in this manner ensured that unbiased test embeddings were generated for every subject out of the entire original dataset so they could be subsequently aggregated to form the fusion dataset.


\subsubsection{Tree-Based Classifier Training} 
To fuse EKYT similarity scores with continuous gaze offset similarity scores, we employ a tree-based score-level fusion approach. 
The EKYT model is frozen, with test embeddings pre-computed before fusion, such that the tree-based model operates exclusively on similarity scores derived from fixed EKYT test embeddings and continuous gaze offset statistics. 
This design prevents any end-to-end interaction between EKYT feature learning and the fusion model. 

To form the dataset that the tree-based model uses for training and testing, the testing set for each EKYT model was used to create authentication and enrollment pairs within the testing set. 
These pairs were then evaluated using the EKYT test embeddings, and, as a result, an unbiased similarity score is produced. This was done for each of the 5 EKYT models. 
The same pairs were then evaluated using the continuous gaze offset method alone to form the continuous gaze offset similarity scores for the pairs. 

Fusion features consist of the raw EKYT and continuous gaze offset similarity scores, along with engineered interaction terms to capture non-linear relationships between the two modalities. 
The tree-based model, therefore, models score-level interactions rather than raw gaze signals, enabling it to learn complementary relationships between temporal biometric similarity and physical continuous gaze offset information. 

To prevent subject identity leakage, the constructed dataset of similarity scores is evaluated using a rigorous 4-fold subject disjoint cross-validation approach. 
For each fold, pairs of subjects were strictly separated such that no subject appeared simultaneously in both the training and testing sets. 
Class imbalance is addressed using sample weighting during training. The final reported metrics represent the average evaluation metric across the 4 folds. 

Multiple tree-based models, specifically Random Forest, Extra Trees, and Gradient Boosting algorithms, were evaluated. 
Model hyperparameters were selected using randomized search exclusively on the training set. 
For testing, the predictions of the individual classifiers were evaluated, and the configuration achieving the lowest EER on the testing set was recorded to determine the optimal fusion strategy. 

\begin{table*}[t]
\centering
\caption{Full comparison of EKYT baseline, tree-based fusion, and weighted fusion across all sequence lengths. Cross-task and Triple fusion are only reported for TEX as they use both RAN and TEX information. Cells highlighted in yellow indicate the lowest value within each row for the corresponding metric.}
\label{tab:full_summary}
\resizebox{\textwidth}{!}{
\begin{tabular}{llc ccccc c ccccc}
\toprule
Dataset & Task & $n_{\text{seq}}$ &
\multicolumn{5}{c}{EER (\%)} &
&
\multicolumn{5}{c}{FRR (\%)} \\
\cmidrule(lr){4-8} \cmidrule(lr){10-14}
 &  &  & EKYT & Tree & Weighted & Cross-task & Triple & & EKYT & Tree & Weighted & Cross-task & Triple \\
\midrule
\multirow{12}{*}{GazeBase}
& \multirow{6}{*}{RAN}
& 1 &  3.7 &  3.5 &  3.7 & -- & -- & & 65.5 & 78.0 &  64.3 & -- & -- \\
& & 2 &  2.2 &  2.2 &  2.2 & -- & -- & &  32.3 & 57.5 & 32.3 & -- & -- \\
& & 3 &  2.2 &  1.8 &  1.9 & -- & -- & & 32.0 & 47.6 &  29.8 & -- & -- \\
& & 4 &  1.4 &  1.3 &  1.4 & -- & -- & &  21.4 & 29.8 & 21.4 & -- & -- \\
& & 6 &  1.3 &  1.2 &  1.3 & -- & -- & & 15.2 & 24.6 & \cellcolor{yellow!50} 14.9 & -- & -- \\
& & 8 &  1.6 &  0.9 &  1.6 & -- & -- & &  10.3 & 24.1 & 10.3 & -- & -- \\
\cmidrule(lr){2-14}
& \multirow{6}{*}{TEX}
& 1 &  1.9 &  2.2 &  1.6 & \cellcolor{yellow!50} 1.2 &  1.5 & & 40.3 & 44.7 & \cellcolor{yellow!50} 40.0 &  51.0 &  45.6 \\
& & 2 &  1.2 &  1.1 &  1.0 &  1.1 & \cellcolor{yellow!50} 0.9 & & 15.9 & 23.8 & \cellcolor{yellow!50} 15.3 &  31.2 &  22.8 \\
& & 3 &  1.2 &  1.2 &  0.9 & \cellcolor{yellow!50} 0.7 &  0.7 & & \cellcolor{yellow!50} 15.2 & 31.3 & 25.8 &  26.2 &  21.5 \\
& & 4 &  0.9 &  1.1 &  0.8 & \cellcolor{yellow!50} 0.5 &  0.6 & &  9.3 & 28.1 & \cellcolor{yellow!50} 9.0 &  23.9 &  11.6 \\
& & 6 &  0.6 &  0.6 &  0.3 & \cellcolor{yellow!50} 0.2 &  0.4 & & \cellcolor{yellow!50}  6.2 & 11.9 &  9.0 &  23.3 &  23.6 \\
& & 8 &  0.5 &  0.5 & \cellcolor{yellow!50} 0.3 &  0.4 &  0.4 & &  4.4 & 14.0 & \cellcolor{yellow!50} 3.4 &  9.3 &  7.8 \\
\midrule
\multirow{12}{*}{GazeBaseVR}
& \multirow{6}{*}{RAN}
& 1 & 18.2 &  17.9 & 18.2 & -- & -- & &  97.8 & 99.5 & 97.8 & -- & -- \\
& & 2 & 12.8 &  12.0 & 12.5 & -- & -- & & 95.6 &  94.8 & 94.8 & -- & -- \\
& & 3 & 10.8 &  10.3 & 10.8 & -- & -- & &  88.9 & 92.4 & 88.9 & -- & -- \\
& & 4 & 10.1 &   9.4 & 10.1 & -- & -- & &  83.1 & 86.5 & 83.1 & -- & -- \\
& & 6 &  8.7 &   8.5 &  8.6 & -- & -- & &  73.7 & 85.5 & 77.6 & -- & -- \\
& & 8 &  8.1 &  8.2 &   8.1 & -- & -- & &  73.7 & 83.3 & 73.7 & -- & -- \\
\cmidrule(lr){2-14}
& \multirow{6}{*}{TEX}
& 1 & 14.6 &  13.6 & 13.9 &  11.7 & \cellcolor{yellow!50} 11.4 & & \cellcolor{yellow!50} 91.1 & 95.0 &  94.3 &  94.1 &  98.2 \\
& & 2 & 11.4 &  10.6 & 10.7 & \cellcolor{yellow!50} 7.2 &  7.4 & & \cellcolor{yellow!50} 77.7 & 84.8 & 90.8 &  93.8 &  94.1 \\
& & 3 &  8.4 &  8.5 &  8.4 &  5.7 & \cellcolor{yellow!50} 5.5 & & \cellcolor{yellow!50} 63.5 & 85.6 & 63.5 &  78.3 &  81.4 \\
& & 4 &  7.1 &   6.9 &  7.1 & \cellcolor{yellow!50} 5.8 &  5.9 & & \cellcolor{yellow!50} 61.7 & 72.4 & 61.7 &  74.3 &  90.7 \\
& & 6 &  5.7 &   5.4 &  5.7 &  5.6 & \cellcolor{yellow!50} 4.8 & & 54.6 & 63.3 & \cellcolor{yellow!50} 51.6 &  55.3 &  77.2 \\
& & 8 &  4.9 & \cellcolor{yellow!50} 4.7 &  4.9 &  5.3 & \cellcolor{yellow!50} 4.7 & & \cellcolor{yellow!50} 49.9 & 65.3 & \cellcolor{yellow!50} 49.9 &  65.0 &  78.1 \\
\bottomrule
\end{tabular}
}
\end{table*}

\subsubsection{Cross-Task and Triple Fusion RF Training}
To fuse the cross-task and triple scores, a simple random forest model is used. This model follows the same dataset formation, splitting, and training procedure as the tree-based classifiers. 
For triple fusion, the engineered feature set is larger because it captures non-linear interactions among 3 scores rather than 2. 
The random forest uses a 4-fold cross-validation approach, reporting the average of the evaluation metric across the 4 folds. 

\subsection{Evaluation and Metrics}
Biometric performance was assessed at 5 – 40 seconds of data in steps of 5 seconds. 
All evaluations were done using both sessions of round 1 data, where session 1 is authentication and session 2 is enrollment. 
To generate the similarity scores for each EKYT model, a 128-dimensional embedding was computed for each fold and concatenated to create a single 512-dimensional embedding for each window. 
The ensemble is then used to predict similarity scores between genuine and impostor samples, which are used to compute the necessary metrics for the baseline. 

We used two metrics to measure and compare the performance of the models. 
These are EER and false rejection rate (FRR) at a fixed false acceptance rate (FAR). 

EER offers a single value for the overall error rate in an enrollment and verification scenario, making it easy to compare the biometric performance of systems. 
To calculate EER, an enrollment and verification set was formed and compared to determine if the model believed they were a match and if the model was correct. 

FRR indicates the feasibility of an EMB system for real-world use. 
FRR tells us how often legitimate users are incorrectly rejected, whereas FAR presents how much of a security risk is posed by impostors gaining access. 
According to FIDO biometric requirements, FRR is used to measure biometric performance and should be no more than 5\% @ $\text{FAR}=10^{-4}$ \cite{FIDOReqs}.  

\section{Results}

This section presents the performance of the EKYT baseline, tree-based fusion, and weighted score fusion across the GazeBase and GazeBaseVR datasets. 
Performance is evaluated using EER. 
Each subsequence corresponds to 5 seconds of gaze data, such that the total duration equals $5 \times n_{\text{seq}}$ seconds.

Table~\ref{tab:full_summary} presents the EER for the baseline and all fusion methods. 
There were no cases in which the baseline outperformed all of the fusion results. 
The lowest achieved EER was 0.2\% by cross-task fusion at $n_{\text{seq}}=6$ on GazeBase data. 
In general, EER is lower when using GazeBase data vs. when using GazeBaseVR data. 

\subsection{Tree-Based Fusion Results}

Tree-based fusion outperformed the EKYT baseline in 17 of the evaluated configurations. 
On GazeBase, it achieved a minimum EER of 0.5\% using TEX task data, demonstrating that a structured combination of modalities can yield strong performance under high-quality recording conditions. 
On GazeBaseVR, it achieved a minimum EER of 4.7\% on the same task, matching or surpassing all other fusion methods and tying only with triple task fusion. 
Both minimum EER values were obtained using 40 seconds of gaze data, suggesting that sufficient signal duration is important for maximizing the benefit of tree-based score combination.

\subsection{Weighted Fusion Results}

Weighted score fusion demonstrated the most consistent improvements over the EKYT baseline, outperforming it in 20 of the evaluated configurations. 
On GazeBase, the best performance was observed on the TEX task, reaching a minimum EER of 0.3\% at both $n_{\text{seq}}=6$ and $n_{\text{seq}}=8$, indicating that a weighted combination of temporal and gaze offset scores is particularly effective under high-quality recording conditions. 
On the RAN task, improvements were more modest, with the lowest EER of 1.3\% at $n_{\text{seq}}=6$.
On the GazeBaseVR dataset, weighted fusion achieved a minimum EER of 4.9\% on the TEX task at $n_{\text{seq}}=8$, matching the baseline in this case.
This suggests that a simple weighted combination may be insufficient to extract complementary information from the noisier VR signal, where more structured fusion approaches such as tree-based or triple fusion, proved more effective.

\subsection{Cross-Task Fusion Results}

Cross-task fusion provided the most consistent improvements across all methods, outperforming the baseline in 33 of the evaluated configurations. 
It also achieved the overall lowest EER observed in this study, reaching 0.2\% on the GazeBase dataset with 30~seconds of gaze data. 
On GazeBaseVR, Cross-task fusion delivered substantial improvements, achieving a minimum EER of 5.3\% at $n_{\text{seq}}=8$. 
Although weighted fusion improved performance, the $\alpha$ values are consistently small, implying that the continuous gaze offset signal contributes marginally to the fused score.

\subsection{Triple-Task Fusion Results}

Triple-task fusion demonstrated strong performance, outperforming the baseline in 31 cases. 
On GazeBase, it achieved a minimum EER of 0.4\% at $n_{\text{seq}}=8$, matching the best cross-task results at that sequence length.
For GazeBaseVR, triple-task fusion achieved its best performance at $n_{\text{seq}}=8$ with an EER of 4.7\%, tying with tree-based fusion and slightly outperforming cross-task fusion.

\section{Discussion}

\subsection{Fusion Methods Compared}
Across both datasets, all three ensemble approaches—Tree-Based, Cross-Task, and Triple fusion—demonstrated consistent improvements over the baseline. 
In contrast, weighted fusion provided mixed results. Because weighted fusion relies on purely linear combinations, it struggles to adapt to interactions between the temporal EKYT signals and continuous gaze offset. 
It proved to be effective during the TEX task on high-quality GazeBase data, achieving 0.3\% EER, but failed to adapt meaningfully to the noisier data introduced by GazeBaseVR.

In contrast, the ensemble methods utilized explicit, non-linear interaction features which allowed the decision trees to adaptively model complex boundaries—emphasizing continuous gaze offset when the temporal signal degraded, and suppressing it when the temporal signal was stronger.

When comparing the ensemble methods among themselves, cross-task fusion tends to perform better on GazeBase data, whereas Triple fusion performs better on GazeBaseVR. 
Since cross-task fusion merges the RAN and TEX tasks, its performance on GazeBase implies that combining only those two signals is optimal when the signal quality is very high. 
In contrast, Triple fusion's performance on GazeBaseVR supports the hypothesis that a physical continuous gaze offset signal provides a complementary signal to EKYT that improves biometric performance, especially in cases where signal quality is degraded, like in the GazeBaseVR dataset.

\subsection{Dataset-Specific Fusion Behavior}

Fusion behavior differs substantially across datasets and tasks, reflecting differences in signal quality, acquisition conditions, and the degree of complementary information available to the fusion model.

On the high-quality GazeBase dataset, the EKYT baseline is already exceptionally accurate (achieving EERs below 1.5\%). Because the baseline signal is so strong, introducing the continuous gaze offset signal provided moderate performance gains. 
The most substantial improvements on GazeBase were achieved through cross-task fusion, which achieved a minimum EER of 0.2\%. 
This indicates that, when using high signal quality data, combining temporal dynamics across tasks yields higher performance gains than adding auxiliary spatial signals.

In contrast, evaluation on the GazeBaseVR dataset demonstrates the values of continuous gaze offset signals under degraded signal quality. 
Baseline EKYT EER values are significantly higher due to headset motion and lower sampling rates. 
In this noisier environment, relying solely on temporal EKYT signals isn't sufficient. 
Continuous gaze offset is inherently tied to spatial accuracy, as it measures the angular error between estimated gaze and intended target locations. In high-quality datasets such as GazeBase, spatial accuracy is consistently high and less variable which limits the contribution of continuous gaze offset. However, in lower-quality datasets such as GazeBaseVR, spatial accuracy is lower which makes continuous gaze offset a more informative biometric signal. 
Employing continuous gaze offset, especially when using non-linear fusion methods, resulted in significant performance boosts. 
Notably, Triple fusion reduced the EER at 5 seconds of data from 14.6\% on the TEX task to 11.4\%.

\subsection{Limitations and Future Work}

This study has several limitations that should be considered when interpreting the results. 
First, fusion was evaluated using simple score-level approaches: linear weighted fusion, tree-based non-linear fusion, and random forest fusion across tasks. 
While non-linear fusion is effective in modeling basic interactions between scores, more sophisticated fusion strategies were not explored. 
More sophisticated methods may better capture complex relationships between temporal biometric similarity and continuous gaze offset, particularly under challenging acquisition conditions.
Second, the experiment evaluation was limited to two datasets, GazeBase and GazeBaseVR. 
While these datasets differ substantially in acquisition environment and signal quality, results derived from only two datasets limit the generalizability of the findings. 
As such, the conclusions drawn in this work should be interpreted as supporting a hypothesis rather than establishing a claim about the effectiveness of continuous gaze offset fusion across eye-tracking systems. 
Future work should therefore focus on validating these findings across additional datasets to assess robustness and generalizability. 

\section{Conclusion}

This study investigated whether continuous gaze offset can complement temporal eye movement biometrics within deep learning frameworks. 
Using the EKYT network as a fixed baseline, we evaluated score-level fusion methods on two datasets: GazeBase (high-quality lab tracker) and GazeBaseVR (lower-quality VR headset). 
Results show the optimal fusion strategy depends on signal quality. In high-quality conditions, temporal baselines already perform well, with cross-task fusion offering the most benefit. 
In degraded VR conditions, adding gaze offset significantly improved performance, especially through triple fusion, combining cross-task temporal dynamics with gaze offset. 
These findings suggest continuous gaze offset serves as a valuable auxiliary signal in challenging environments, and since improvements came via score-level fusion, existing pipelines can be enhanced without retraining underlying models.

\section*{Privacy and Ethics Statement}

This study utilizes anonymized, publicly available eye-tracking data.
Only gaze data is used, and we strongly advocate for the careful and transparent application of these techniques to prevent potential misuse and to promote user trust in future eye-tracking technologies.

\bibliographystyle{unsrt}  
\bibliography{01ms}

\end{document}